# All-solid-state VUV frequency comb at 160 nm using multi-harmonic generation in a non-linear femtosecond enhancement cavity

J. Seres,[1,][*] E. Seres,[1] C. Serrat,[2] Erin C. Young,[3] James S. Speck,[3] T. Schumm,[1]

[1] Atominstitut - E141, Technische Universität Wien, Stadionallee 2, 1020 Vienna, Austria

[2] Universitat Politècnica de Catalunya, Departament de Física, Colom 11, 08222 Terrassa, Spain

[3] Materials Department, University of California, Santa Barbara, CA 93106-5050, USA

We report on the realization of a solid-state-based vacuum ultraviolet frequency comb, using multi-harmonic generation in an external enhancement cavity. Optical conversions in such arrangements were so-far reported only using gaseous media. We present a theory that allows selecting the most suited solid generation medium for specific target harmonics by adapting the bandgap of the material. Consequently, we experimentally use a thin AlN film grown on a sapphire substrate to realize a compact frequency comb multi-harmonic source in the DUV/VUV spectral range. Extending our earlier VUV source [Opt. Exp. 26, 21900 (2018)] with the enhancement cavity, a sub-Watt level Ti:sapphire femtosecond frequency comb is enhanced to 24 W stored average power, its 3$^{rd}$, 5$^{th}$ and 7$^{th}$ harmonics are generated, and the target harmonic power at 160 nm increased by two orders of magnitude. The emerging non-linear effects in the solid medium together with suitable intra-cavity dispersion management support optimal enhancement and stable locking. To demonstrate the spectroscopic ability of the realized frequency comb, we report on the beat measurement between the 3$^{rd}$ harmonic beam and a 266 nm CW laser reaching about 1 MHz accuracy.

## I. INTRODUCTION

The extension of frequency comb metrology beyond the UV spectral range into the DUV [1-4] and even into the VUV and XUV [5-9] gives an opportunity to measure new important atomic and molecular transitions for testing quantum electrodynamics or to look for new atomic clock transitions. Perturbative non-linear conversion methods in solids such as second- and sum-frequency generation are limited at about 160 nm by the transparency of the used non-linear crystal [10-12], and have been realized at low (kHz or lower) repetition rates. To reach short wavelengths at high repetition rates, non-perturbative high-harmonic generation is currently the only candidate. Such high-harmonic sources were successfully realized by adding an enhancement cavity to fiber laser systems or Ti:sapphire oscillators [13-18]. In these realizations, different noble gases were used for non-perturbative frequency conversion, which relies on the ionization of the gases and hence requires suitably high (> 10$^{13}$ W/cm$^2$) laser peak intensity. Such intensity cannot be maintained easily at the very high repetition rate (typically > 100 MHz) of the frequency combs, especially those based on Ti:sapphire oscillators. Another crucial problem at such high repetition rates is that the residual ionized gas plasma produces instabilities within the enhancement cavity [15, 19] which in some cases can be overcome with special techniques like using high temperature mixed gases [20].

In this study, we report the first realization of a VUV frequency comb using solid material as the medium for non-perturbative multi-harmonic generation within an enhancement cavity. (We refer to our observations as "multi-harmonic generation", as only 3$^{rd}$, 5$^{th}$ and 7$^{th}$ orders are observed with the available detectors.) Although very early realizations [21,22] were reported, solids as generation media have recently attracted new attention because they require lower laser peak intensities (> $10^{11}$ W/cm$^2$) than gases for reaching the VUV or XUV spectral ranges and they can operate even at MHz's repetition rates [23-25]. Consequently, using a solid as generation medium is attractive to produce femtosecond pulses at 160 nm (our target wavelength) by directly generating the 5$^{th}$ harmonic of a Ti:sapphire oscillator at 108 MHz repetition rate (800 nm central wavelength). Intense 5$^{th}$ harmonic is generated from a thin AlN crystalline film grown on a sapphire substrate within a compact setup and the appearance of the 7$^{th}$ harmonic is demonstrated (see Fig. 2).

## II. EXPERIMENTAL SETUP

The optical setup can be seen in Fig. 1. The 0.9 W average power output of a Ti:sapphire oscillator-based frequency comb (FC8004, Menlo Systems), generating 27 fs-long pulses at a repetition rate of 108 MHz, centers at 800 nm central wavelength. The repetition rate and the offset frequency of the frequency comb are locked to a 10-MHz Rubidium frequency standard. The pulses are negatively chirped by chirp-mirror pairs (175 fs$^2$/bounce, 26 bounces) to pre-compensate the dispersion of the optical elements in the optical path before entering the enhancement cavity and a wedge-pair is added for dispersion fine-tuning. Laser pulses with about 7.5 nJ energies (~0.8 W) seeded the cavity through an input coupler (IC) mirror with 2% transparency. The cavity is built inside a vacuum chamber to avoid absorption of the generated VUV signal in air. The group delay dispersions (GDD) of the mirrors are chosen to yield near zero GDD per roundtrip for optimal performance (more details will be given below). The length of the ring-type enhancement cavity is 2776 mm to match the seeding laser repetition rate. One cavity mirror is mounted on a piezo to lock the cavity length to the repetition rate of the Ti:sapphire oscillator, using a Hänsch-Couillaud scheme [26].

Inside the enhancement cavity, two curved mirrors with focal lengths of 50 mm form a focus where the solid medium is positioned for multi-harmonic generation. For that purpose, a 30-nm-thick AlN crystalline film has been grown on a 100-µm-thick sapphire substrate. Details about the preparation of the AlN film are given in the section Appendix and a justification for choosing AlN will be given below. The sample is tilted to near-Brewster angle to ensure minimal loss, which is critical for reaching high enhancement factors. This part of the setup is magnified for better visibility in Fig. 1(b). The cavity alignment and focusing conditions are monitored at the secondary focus by observing leaked signal from one of the cavity mirrors. The generated 5$^{th}$ harmonic beam (central wavelength at 160 nm) is extracted from the cavity using a multilayer mirror designed for >90% reflectivity within the 150-170 nm spectral range [16]. The output coupler mirror is placed far from the focus to avoid damage of the multilayer coating. The extracted 5$^{th}$ harmonic beam together with some residual (~1%) of the 3$^{rd}$ and the 7$^{th}$ harmonics is focused to the input slit of a VUV spectrometer (McPherson 234/302), equipped with a 300 lines/mm grating, using a VUV-grade MgF$_2$ lens. According to earlier measurements [23], the harmonic beams co-propagate with the fundamental laser beam; any small deviation that might still exist is corrected by the lens that collects the harmonics onto the spectrometer slit. The cavity and the VUV spectrometer are in vacuum with a background pressure of 10$^{-3}$ mbar. In certain measurements, a VUV bandpass filter was inserted into the harmonic beam at the entrance of the spectrometer to suppress the 3$^{rd}$ harmonic even further. The spectrally dispersed beam behind the output slit was detected with a VUV photomultiplier (Hamamatsu R6836), sensitive in the 115-320 nm spectral range. The offset frequency of the oscillator is adjusted manually; this adjustment is critical for reaching optimal harmonic power.

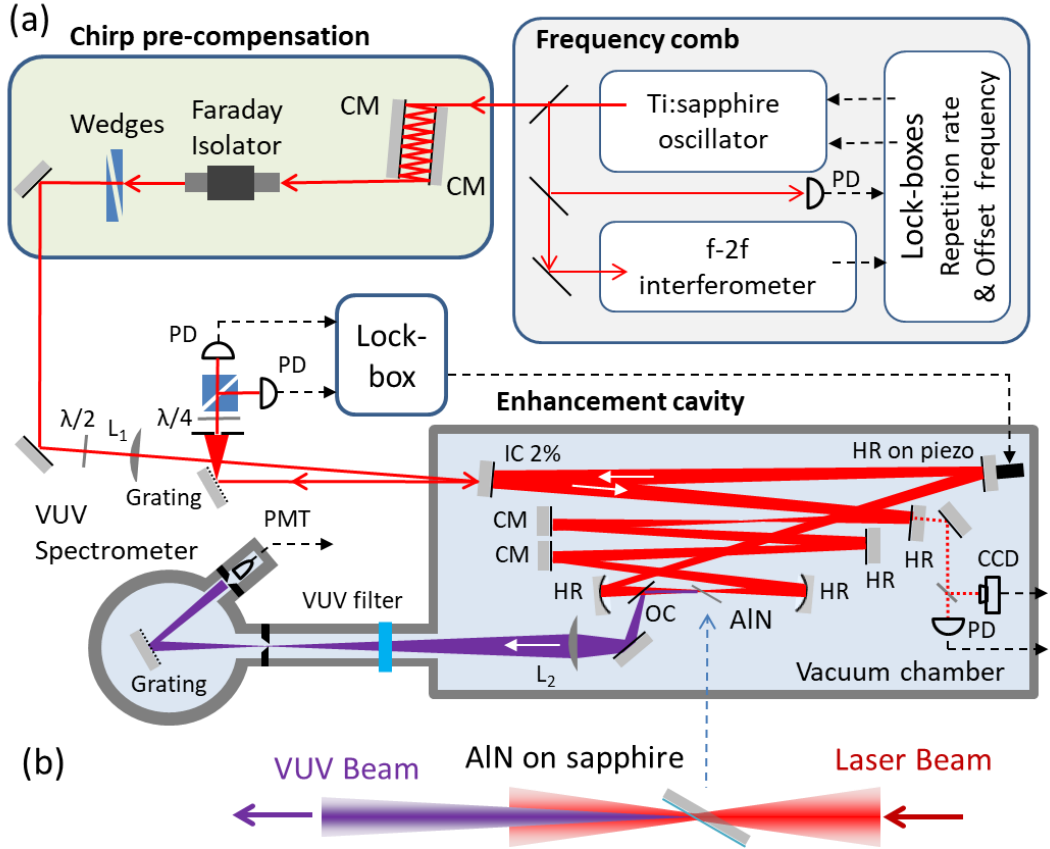

FIG. 1. Experimental setup with (a) HR: high reflectors; CM: chirped mirror; $L_1$: BK7 lens, f = 750 mm; $L_2$: $MgF_2$ lens, f = 160 mm; IC: input coupler, OC: output coupler; PBS: polarizing beam splitter; PD: photo diodes, PMT: photo-multiplier tube. (b) Magnified area around the crystal.

**III. OPTIMIZING INTRA-CAVITY DISPERSION**

The substrates of both, the output coupler mirror and the AlN film introduce some (approx. +16 $fs^2$) positive GDD's in the enhancement cavity, which is compensated by a combination of broadband zero-dispersion high reflectors (HR) and chirped mirrors (CM) with negative GDDs of -10 $fs^2$ and -40 $fs^2$ (Layertec) built into the cavity, see Fig. 1(a). To find the optimal conditions for efficient $5^{th}$ harmonic generation, a measurement series was performed by introducing specific negative GDD between 0 $fs^2$ and -50 $fs^2$, the results are plotted in Fig. 2. As shown in Fig. 2(a), by far the strongest harmonic signal can be obtained for a GDD of -40 $fs^2$, when the intensity of the $3^{rd}$ and $5^{th}$ harmonics are about 50-times higher than for any other GDD and even the $7^{th}$ harmonic at 114 nm appears. The comparison of the powers of the intra-cavity laser and its $3^{rd}$ and $5^{th}$ harmonics are depicted separately in Fig. 2(b). In this figure, the absolute detected harmonic powers are plotted, which are the signals within the respective harmonic line that reached the photomultiplier, corrected by the known spectral sensitivity of the photomultiplier and the diffraction efficiency of the grating. These values are indicative lower bounds, as we were not able to reliably determine which fraction of the generated signal passed the output coupler, the steering mirror, the focusing lens and passed through the input slit of the spectrometer.

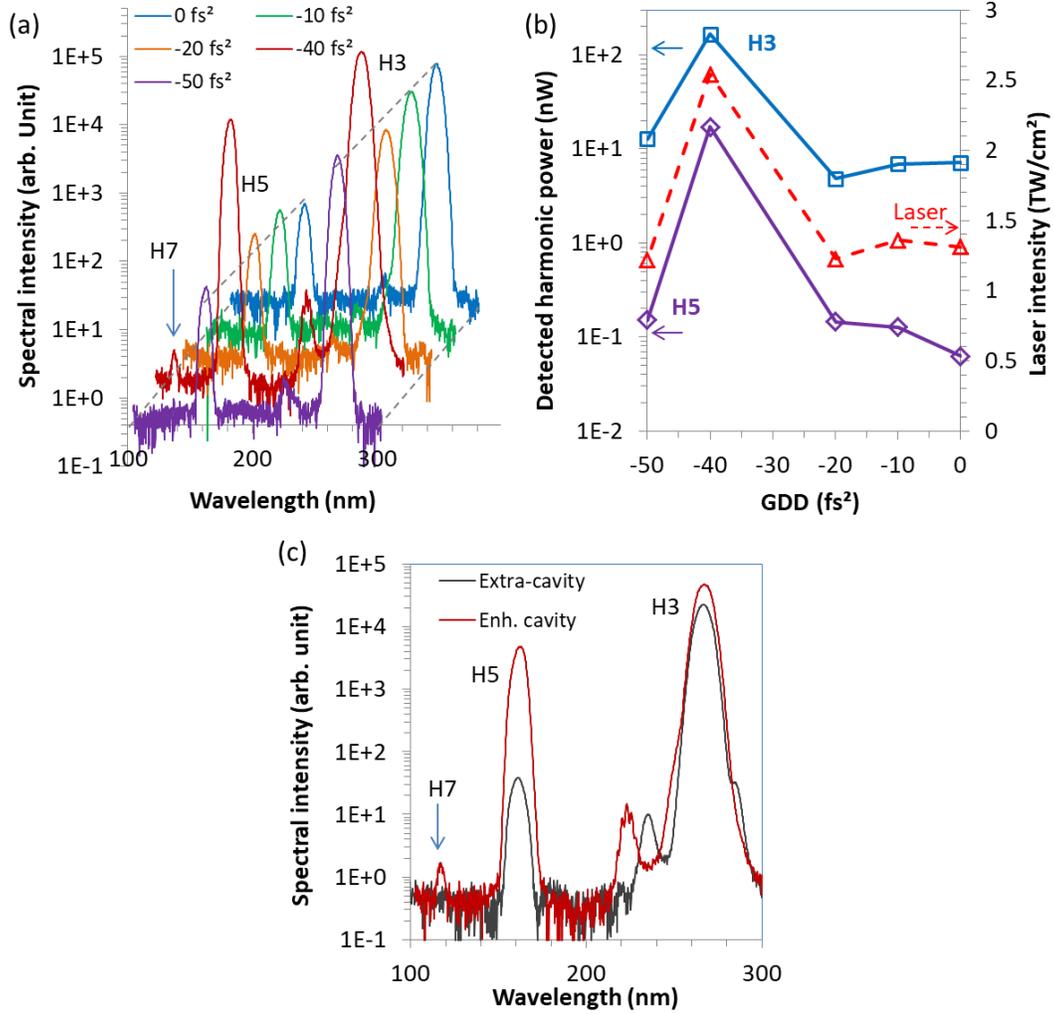

FIG. 2. (a) Generated multi-harmonic spectra at several introduced intra-cavity GDDs. The dashed grey lines help to guide the eyes at the base lines and at the harmonic orders. (b) The power of the generated harmonics is strongly affected by the introduced intra-cavity GDD and follows the peak intensity of the intra-cavity laser pulses. (c) Comparing to an extra-cavity setup, when the harmonics are directly generated with the output of the Ti:sapphire oscillator [25], strong increase of the 5$^{th}$ harmonic power is obtained by placing the AlN sample into the enhancement cavity.

The obtained powers of the generated harmonics are dramatically increased by placing the AlN sample into the enhancement cavity. In Fig. 2(c), we compare the spectrum generated in the enhancement cavity (red curve) to the one generated in an extra-cavity setup [25] (grey curve) using the same AlN sample. In comparison with the extra-cavity arrangement, we successfully increased the pump peak intensity of 1±0.1 TW/cm$^2$ at average power of ~0.8 W to 2.5±0.2 TW/cm$^2$ at average power of ~24 W, and the generated harmonic power at 160 nm was enhanced by a factor of ~120. Note that the real generated intensities of the 3$^{rd}$ and 7$^{th}$ harmonics should be about 100-times higher than they appear in the spectra obtained from the enhancement cavity because of the low (~1%) reflectivity of the output coupler at those harmonic wavelengths.

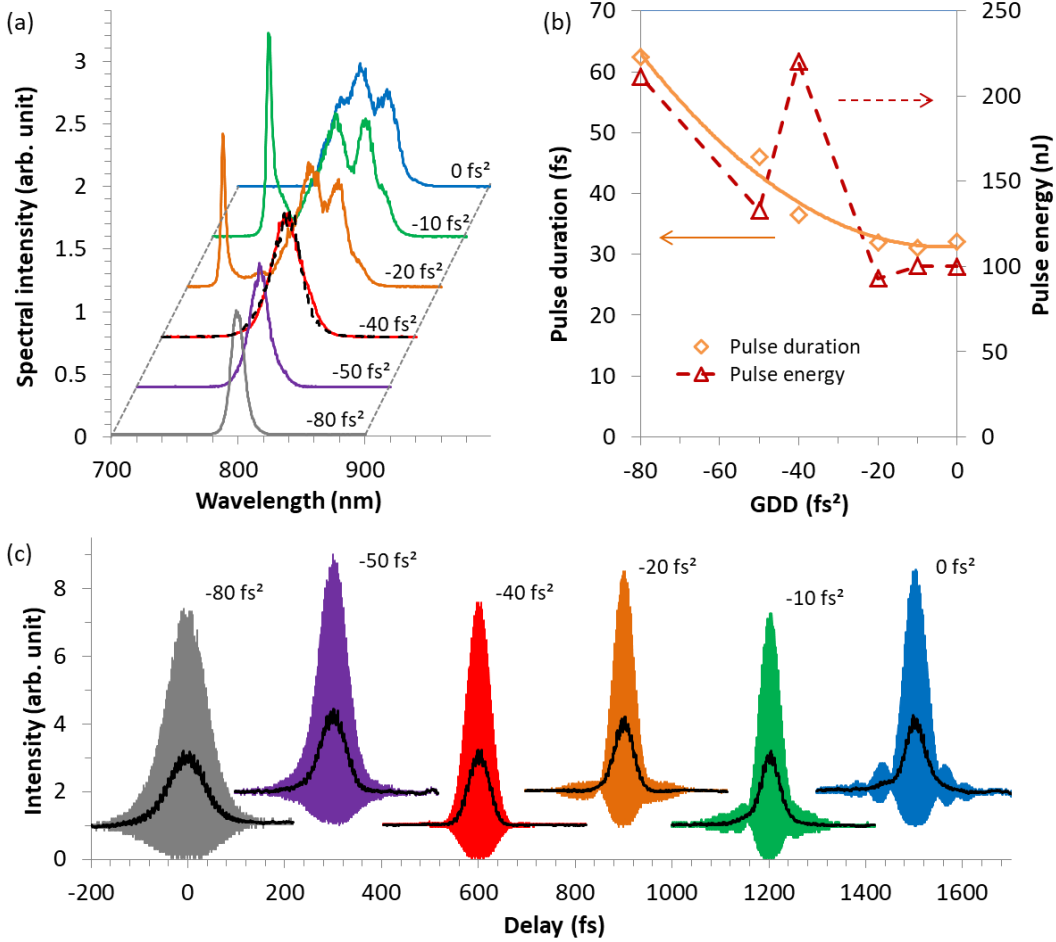

FIG. 3. (a) The intra-cavity spectrum, (b) the intra-cavity pulse energy and pulse duration are dependent on the introduced intra-cavity GDD. At the GDD of -40 fs$^2$, the intra-cavity spectrum almost perfectly overlaps with the seed spectrum (black dashed). (c) Measured autocorrelation curves of the intra-cavity pluses.

The reason for the much higher harmonic yield at -40 fs$^2$ GDD can be understood by inspecting the measured intra-cavity laser spectra in Fig. 3(a) and the measured autocorrelation curves of the intra-cavity laser pulses in Fig. 3(c). The intra-cavity spectrum at -40 fs$^2$ is almost perfectly matching the spectrum of the oscillator (black dashed line) used to seed the cavity. This yields the optimal conditions for building up the highest power, enhancement rate, and pulse energy in the cavity as can be observed in Fig. 2(b) and 3(b). Additionally, this GDD produces the most clean pulse shape without side wings as can be seen in Fig. 3(c). These two factors together support an exceptional high yield of harmonics even if the pulse duration is not the shortest one in this case, see Fig. 3(b). The intra-cavity pulse energy here reaches 220 nJ, which indicates an enhancement factor of about 30 when comparing with the about 7.5 nJ energy of the seed pulse.

## IV. NON-LINEAR ENHANCEMENT CAVITY

Intuitively, one would expect optimal conditions for harmonic generation at a zero net intra-cavity GDD (added intra-cavity GDD between -10 and -20 fs$^2$). Non-linear effects in the source (AlN film and the sapphire substrate) however alter this behavior significantly. Close to the near-zero net GDD, the spectrum in the cavity gets broader than the seed spectrum, see Fig. 3(a). The spectral components outside of the seed spectrum cannot build up and experience only losses. This limits the storable intra-cavity power and consequently the harmonic yield. The detrimental effect of the strong non-linearity can also be observed in the cavity-length detuning curve (orange lines) plotted in Fig. 4(a) where larger scanning time means shorter cavity length. In the linear case, the optimal condition for the maximum intra-cavity power should be at zero detuning. The non-linear effect however shifts the maximum to a shorter cavity length and the stored power drops, forming a shoulder [indicated by an arrow in Fig. 4(a)] at the falling edge of the curve. The 5$^{th}$ harmonic is generated at this shoulder at reduced intra-cavity power.

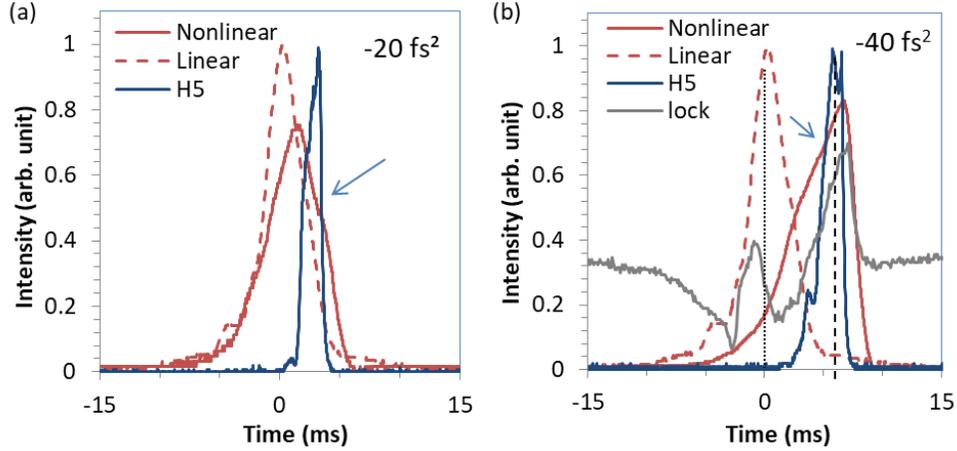

FIG. 4. Effect of the non-linearity on the cavity-length detuning curves at added intra-cavity GDD of (a) -20 fs$^2$, (b) -40 fs$^2$, (orange solid/dashed: intra-cavity laser intensity; blue: 5$^{th}$ harmonic intensity H5; grey: signal from the locking electronics). The locking possibilities in panel (b) are indicated in the linear (dotted black) and the non-linear (dashed black) cases.

In the case of -40 fs$^2$ added GDD, however, the built-up spectrum should have been narrower than that of the seed (almost as for -50 fs$^2$) but the non-linear effect broadens the spectrum. In this case, the spectral narrowing due to non-optimal GDD is fully balanced by the non-linear spectral broadening, giving optimal conditions for building up the intra-cavity power and for higher yield of harmonics. This positive effect can be seen also in the cavity detuning curve (solid lines) of Fig. 4(b). At about the scanning time of 4-5 ms, a shoulder also appears (denoted with an arrow), however, because of the better build-up condition, the stored intra-cavity power increases here instead of decreasing and produces a strong harmonic signal (blue, H5).

Although non-linear effects in an enhancement cavity generally deteriorate the conditions for locking the cavity [19, 27-30], we successfully modified the Hänsch-Couillaud scheme to reach locking at near-optimal harmonic generation conditions. It required only the selection of the wavelength used for producing the error signal at few-nm below the central 800 nm (optimal in linear case), and inverting the gain of the locking electronics. The locking signal (grey) obtained during a cavity length scan is plotted in 4(b). In linear mode, the intra-cavity power (orange dashed line) reaches its maximum at zero detuning and the cavity can be locked on the falling slope of the locking signal (grey) denoted with black dotted line. In non-linear mode, the cavity can also be locked at zero detuning to the falling slope of the locking signal, however, then the intra-cavity power is low and there is no generated 5$^{th}$ harmonic (or very weak). By inverting the gain of the lock electronics, the non-linear cavity can be locked to the rising slope (denoted with black dashed line) when the harmonic yield is at or near to its optimum.

## V. PRESERVATION OF THE FREQUENCY COMB STRUCTURE

To ensure that the generated harmonics preserve the frequency comb structure of the seed laser, a test measurement was performed and beat signals were generated between the 3$^{rd}$ harmonic beam and a UV CW laser (Toptica TopWave 266) having spectral bandwidth < 1 MHz. The measurement was realized with the 3$^{rd}$ harmonic similarly to Ref. [13] because suitable CW lasers for the 5$^{th}$ harmonic at around 160 nm are currently not available. The experimental setup is plotted in Fig. 5(a). The original setup of Fig. 1(a) was slightly modified by inserting a MgF$_2$ window into the harmonic beam, used to combine the beams of the two sources. A lens telescope (1:4) of L$_1$ and L$_2$ adapted the beam size of the CW laser to the harmonic beam and a lens L$_3$ focused the beams to the input slit of the spectrometer. The measured spectra of the harmonic beam and the CW laser are plotted in the inset of Fig. 5(a). In the case of the CW laser, the width of the line is limited by the resolution of the spectrometer. The Fourier spectrum of the time dependent beat signal measured by the photo-multiplier is plotted in Fig. 5(b). It shows the 108 MHz peak at the repetition rate of the oscillator (distance between the comb lines for all harmonic orders) and the two beating peaks formed by the CW laser with the nearest two comb lines. Because the spectral line of the CW laser is located near two comb lines, which are ~10$^7$ multiples of the repetition rate, a sub-Hz tuning of the repetition rate of the seed comb results in an observable tuning of the beat signal, which can be seen in the inset of Fig. 5(b) for two additional measured repetition rate settings. The comb structure of the generated multi harmonics is hence clearly demonstrated.

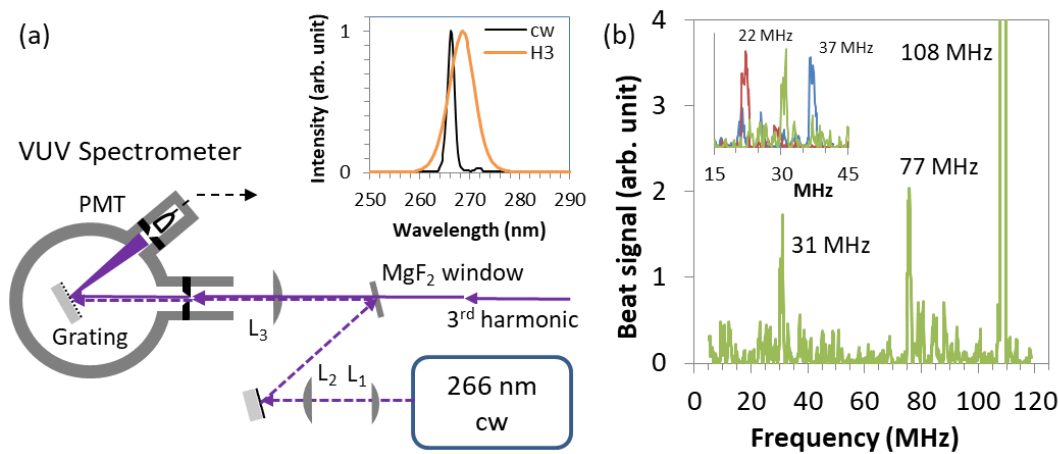

FIG. 5. Beat measurement to verify the frequency comb character of the generated harmonic beam. (a) Experimental setup. Inset: measured spectra of the 3$^{rd}$ harmonic beam and the CW laser. (b) Beat-signal peaks relative to the two nearest comb lines (distance 108 MHz) are observed and (inset) tuned by sub-Hz tuning of the repetition rate of the seed laser.

## VI. BANDGAP TUNING: USING ALN FILM AS MHG SOURCE

We chose AlN crystalline film to generate strong 5$^{th}$ order harmonics at 160 nm. AlN is III-V type wide-bandgap (6.3 eV) semiconductor with a high damage threshold of ≈100 TW/cm$^2$ for femtosecond pulses [31, 32], which allows us to apply high intensity laser pulses. Additionally, it can be grown in high quality on sapphire substrates, which has an even higher band gap of 8.7 eV, a ~50 % higher damage threshold [31, 32] and a small non-linear refractive index [33], n$_2$ = 3×10$^{-16}$ cm$^2$/W. In Fig. 6(a), the measured spectrum generated on the back surface of the sapphire substrate alone (without AlN film) is compared to the spectrum obtained when the back surface of the sapphire comprises the AlN film. As reported in previous work [25], harmonics are more intensely generated if the film is situated on the back surface. In the cavity, the -40 fs$^2$ dispersion compensation yields the strongest harmonics, as described above.

However, with the sapphire substrate alone, the same intra-cavity power and perfect spectral overlap cannot be obtained. This reveals that the non-linearity of the AlN film [34] $n_2 = 1\times10^{-12}$ cm$^2$/W also contributes significantly, which can be expected from the similar ($3\times10^{-18}$ cm$^3$/W) $n_2L$ products of the AlN film and the sapphire substrate. Therefore, in the case of AlN, we stabilized the cavity to the same intra-cavity peak intensity of 1.6 TW/cm$^2$, as it was obtained with the substrate alone, for making the comparison. As depicted in Fig. 6(a), when the sapphire substrate comprises the AlN film, the generated 3$^{rd}$ harmonic is only about 3-times stronger than from the substrate alone, but the 5$^{th}$ harmonic is about 1000-times stronger. The 7$^{th}$ harmonic is hardly recognizable at such low laser intensity.

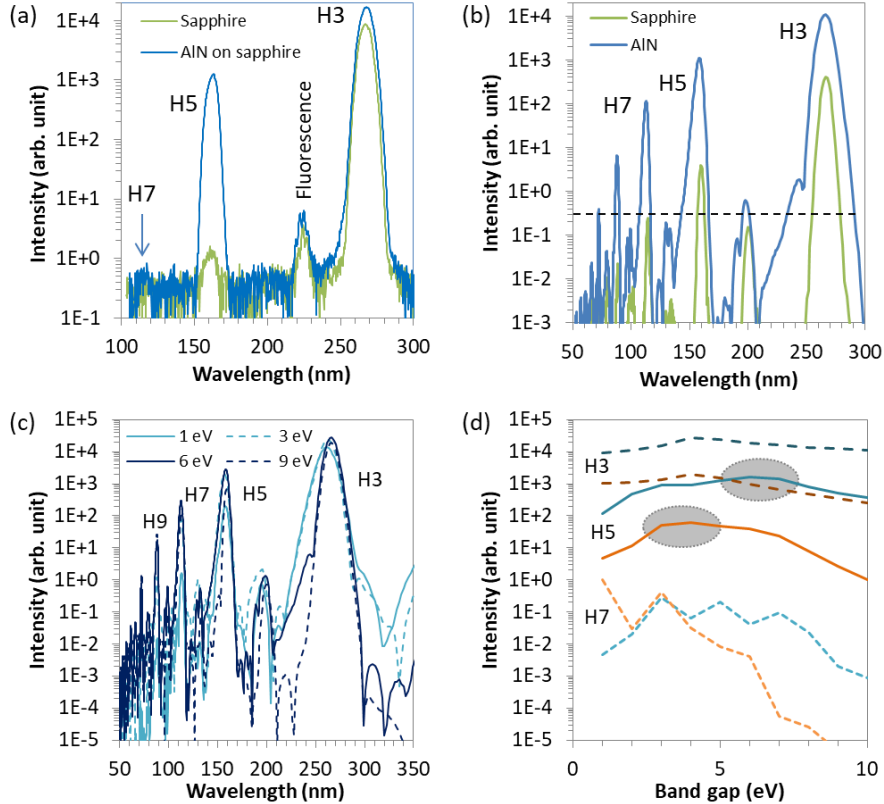

FIG. 6. (a) Measured multi-harmonic spectra generated from an AlN film on sapphire (blue) and from the sapphire substrate without film (green) at laser intensity of 1.6 TW/cm$^2$. The origin of the fluorescence peak at 220 nm is discussed in the text. (b) Calculated spectra of sapphire (green) and AlN (blue) for the same condition as in panel (a), the dashed black line indicates the detection limit. Rabi frequencies $\hbar\Omega_R$ are 2.2 eV and 6.5 eV, respectively. (c) Calculated spectrum series and (d) the intensities of the individual harmonic lines for AlN by changing the band gap energy between 1 eV and 10 eV. Calculations in (d) were performed for H3 (dashed lines), H5 (solid lines) and H7 (fine dashed lines) and for two transition dipoles of $1\times10^{-28}$ Cm (orange) and $3\times10^{-28}$ Cm (blue), or Rabi frequency of 2.2 eV and 6.5 eV, respectively.

Our measurements therefore reveal that, for the present parameters and configuration, AlN on sapphire results in a much preferable configuration for the generation of the 5$^{th}$ harmonic at 160 nm than sapphire alone. The effect of AlN was studied with numerical simulations, which were performed based on the optical Bloch equations as described in Ref. [35]. We modelled multi-harmonic generation by considering the simplest tight-binding one conduction and one valence band geometry. The model calculated the harmonics generated by both interband transitions and intraband currents. We performed calculation series for sapphire alone (band gap $E_g$ = 8.7 eV, reduced effective mass of electron-hole pairs m* = 0.3, and lattice constant $a$ = 0.476 nm) and for AlN ($E_g$ = 6.3 eV, m* = 0.4, and $a$ = 0.310 nm). For AlN, the transition dipole moment at the band gap was estimated from the known absorption coefficient α = 3×10$^5$ cm$^{-1}$, and the effective density of states at the CB edge $N_c$ = 6.2×10$^{18}$ cm$^{-3}$ as $d_{cv}(k=0) \approx \sqrt{\frac{\hbar c \varepsilon_0 \alpha}{2\pi N_c}}$ = 4.6×10$^{-28}$ Cm. No estimation of the dipole moment could be performed for sapphire. We calculated the harmonic spectra, by varying the dipole moments in the simulations and compared them with the measured ones. The driving pulse was considered to be of Gaussian shape in time with the duration of 27 fs and a central wavelength of 800 nm. The dipole decay time is chosen as $T_2$ = 1 fs. The calculated spectra that best fitted the measurements are plotted in Fig. 6(b). While the calculations obviously have a much higher dynamic range, the noise limit of the measurements is indicated by a black dashed line. The best agreement with the corresponding measurements were found for AlN at the dipole moment of 3×10$^{-28}$ Cm, which is very close to the estimated value (4.6×10$^{-28}$ Cm), and 1×10$^{-28}$ Cm for sapphire, which give Rabi frequencies ($\hbar\Omega_R$) as 6.5 eV in the case of AlN and 2.2 eV for sapphire at the peak intensity of 1.6 TW/cm$^2$.

In order to comprehend the different power obtained for the 5$^{th}$ harmonic in the case of AlN+sapphire compared to the case of sapphire alone, calculation series were performed for AlN by varying the band gap energy while keeping all other parameters unchanged, and considering two different Rabi frequencies of 2.2 eV and 6.5 eV. Some of the calculated spectra and the intensity dependence of the 3$^{rd}$, 5$^{th}$ and 7$^{th}$ harmonics are plotted in Figs. 6(c) and 6(d). One can observe that for the 5$^{th}$ harmonic generation, the optimal band gaps are close to (or somewhat larger than) the corresponding Rabi frequencies [shaded areas in Fig. 6(d)]. Although the model that we consider is too simple to quantitatively describe all the physics involved in the measurements, we observe that for the present parameters and configuration the 6.3 eV bandgap of AlN favors an optimal generation of the 5$^{th}$ harmonic, which is not the case for sapphire, seemingly because the 8.7 eV band gap is too high compared to the used 2.2 eV Rabi frequency.

Beyond the 5$^{th}$ and 3$^{rd}$ harmonics, in the spectra in Figs. 2(a) and 6(a), a peak can be recognized at about 220 nm. Although, AlN has a photoluminescence peak at around this wavelength [36] this peak appeared also without the AlN film, as can be seen in Fig. 6(a). The origin of this fluorescence peak is still fully not clarified. We suspect it to be the fluorescence of the residual air in the vacuum chamber ($O_2$, NO or ozone), exited by the 3$^{rd}$ harmonic beam via two-photon absorption.

## VII. CONCLUSION

An all-solid-state DUV/VUV frequency comb was developed by generating 3$^{rd}$, 5$^{th}$, and 7$^{th}$ harmonics of a Ti:sapphire oscillator using AlN thin film within an external enhancement cavity. Introducing solid matter into the focus of the enhancement cavity leads to a noticeable non-linearity. We successfully compensated for this and developed a locking scheme to stabilize the enhancement cavity at optimal harmonic yield. Using numerical simulations, we found that AlN with its 6.3 eV band gap is an almost optimal material for the efficient generation of our target wavelength of 160 nm (5$^{th}$ harmonic). The realized multi-harmonic source can be used for high-precision spectroscopy of transitions in atoms or the $^{229m}$Th isomer nucleus [37, 38] in the DUV/VUV spectral range. Furthermore, because the generated harmonic lines are in the UV-C spectral range, it can be used to study fast disinfection effects and ozone formation. With this first demonstration of an enhanced all-solid-state VUV frequency comb, at 160 nm, we successfully increased the generated harmonic power by two-orders of magnitude compared to an extra-cavity arrangement and reached a detected conversion efficiency of ~3×10$^{-8}$, while remaining at least one order of magnitude below the damage threshold of our multi-harmonic sample. Our concept has hence the capacity to increase the harmonic power and conversion efficiency by orders of magnitude and to extend the spectral range to higher order XUV harmonics, assuming the intra-cavity peak power can be increased and a suitable output coupler can be used.

**ACKNOWLEDGEMENT**

This project has received funding from the European Union's Horizon 2020 research and innovation program under Grant Agreement No. 664732 ("nuClock") and Grant Agreement No. 820404 („iqClock") and was supported by the WWTF Project No. MA16-066 ("SEQUEX"); and by the Spanish Ministry of Economy and Competitiveness through "Plan Nacional" (FIS2017-85526-R). The work at UCSB was supported in part by the KACST-KAUST-UCSB Solid State Lighting Program.

The authors thank TOPTICA Photonics for providing the TopWave 266 nm laser used in the beating measurements.

**APPENDIX: AlN SAMPLE PREPARATION**

As solid medium, thin AlN (0001) films, nominally 30, 100 and 300 nm thick, were grown by molecular beam epitaxy (MBE) in a Veeco Gen 930 MBE system using a solid source for Al and thermally cracked ammonia as active nitrogen source. The Al flux was on the order of $10^{-8}$ Torr and $NH_3$ pressure during growth was approximately $10^{-6}$ Torr. The substrates were 100 μm sapphires and were outgassed at 400 °C for one hour prior to AlN deposition at 760 °C. Reflection high energy electron diffraction during growth of the AlN layers indicated continuous film growth but with rough surface morphology for the 100 and 300 nm films. After testing the different samples, the 30-nm-thick AlN film was found to be most suitable for intra-cavity multi-harmonic generation.